\documentclass[a4paper,11pt]{article}
\usepackage{pos}
\usepackage[utf8]{inputenc} % allow utf-8 input
\usepackage[T1]{fontenc}    % use 8-bit T1 fonts
\usepackage{hyperref}       % hyperlinks
\usepackage{url}            % simple URL typesetting

\title{Higgs criticality in and beyond the Standard Model}

\author*{Thomas Steingasser}

\affiliation{%
Department of Physics, Massachusetts Institute of Technology, Cambridge, MA 02139, USA }
\affiliation{Black Hole Initiative at Harvard University, 20 Garden Street, Cambridge, MA 02138, USA
}%

\emailAdd{tstngssr@mit.edu}

\abstract{The properties of the Higgs potential are determined by three parameters: the mass parameter, the quartic self-coupling, and a constant term. Remarkably, all three of these parameters seem subject to a significant amount of fine-tuning, relating to the hierarchy problem, the metastability of the electroweak vacuum, and the cosmological constant problem. All these tunings can be seen as their corresponding parameters being close to critical values marking quantum phase transitions. While such behavior is surprising from a conventional particle physics perspective, it is a common feature of dynamical systems. This has motivated the conjecture that the values of the Higgs' parameters are the result of some dynamical mechanism. 

This possibility suggests the construction of mechanisms dynamically choosing sets of Higgs parameters. In these notes, I discuss a complementary approach. Taking seriously that such a mechanism could exist in nature, it is plausible to assume that it also influences Beyond-Standard-Model physics. This suggests considering near-criticality in any model of interest and investigating its consequences more generally, in particular independent of a concrete mechanism responsible for it.

I first explain what it means for the parameters of the Higgs potential to be near-critical. This includes a discussion of the recently discovered ``metastability bound'' on the Higgs mass, which can be understood through a critical point.

I then review two concrete examples of mechanisms in which the parameters of the Higgs potential are dynamically driven towards critical values. These mechanisms also serve as an important proof on concept for the feasibility of the assumptions at the foundation of these notes. Using a simple example for concreteness, the final part of these notes then explicitly demonstrates how to approach a given model in the light of the near-criticality conjecture.}

\FullConference{%
  Workshop on the Standard Model and Beyond\\
  05. September 2023\\
  Mon Repos, Corfu, Greece
}

\begin{document}
\maketitle
\section{Introduction}
More than a decade after the discovery of the Higgs boson, its properties persistently evade an explanation in terms of established particle physics principles such as naturalness~\cite{Giudice:2008bi,Giudice:2013yca,Giudice:2017pzm}. Its observed mass of 125 GeV together with the absence of signs of new physics throughout the LHC's entire 14 TeV range lies at the heart of the hierarchy problem. Its couplings meanwhile seem to lie in a small region of parameter space for which radiative corrections to the quartic coupling give rise to a second minimum of its potential at very high energies, whose properties are such that the Higgs can tunnel into it out of its current minimum, the \textit{electroweak vacuum}, but only with a highly suppressed rate. The constant term in the potential can similarly be understood as fine-tuned against the gravitational contribution to the cosmological constant, together giving rise to the cosmological constant problem. What makes this behavior so puzzling from the perspective of conventional particle physics is that it requires a large degree of fine-tuning of apparently unrelated parameters.

This tuning can be understood in terms of so-called \textit{quantum phase transitions} (QPTs). Unlike their finite-temperature analogues, a QPT describes the transition of a system from one state to another through the change of a physical parameter rather than finite-temperature effects~\cite{Jaeger:1998,Carr2010,Sachdev2011}. For the case of a scalar field, this breaks down to quantum corrections to the potential's parameters leading to a change of the qualitative features of the potential, in particular the emergence and disappearence of additional minima. A set of parameters marking such a transition is commonly referred to as \textit{critical}, and the apparent tunings observed in the SM can be understood as \textit{near-criticality}. A small mass term relative to the theory's natural energy scale, for instance, can be understood as being tuned to a value close to the transition from a phase with spontaneous symmetry breaking to one without. Similarly, the running of the quartic coupling to negative values at high energies lies close to the transition from a potential with a unique minimum at low energies to one with a second, lower-lying minimum at high energies. While puzzling in the context of particle physics, such behavior is a common feature of dynamical systems, where it is known as \textit{self-organized criticality}~\cite{PhysRevLett.59.381}. Altogether, this suggests that the parameters of the Higgs potential are neither fundamental constants nor obtained from some UV completion in the usual way, but rather determined through some dynamical mechanism. This possibility is often referred to as \textit{(dynamical) vacuum selection}, as each set of possible parameters can be understood as linked to a distinct vacuum of the underlying theory.

Most importantly for our purpose, this suggests that \textit{some} mechanism might exist in nature which has driven the Higgs potential's parameters to its near-critical values. This immediately suggests to actually develop such a mechanism - and indeed, several authors have pursued this endeavor~\cite{Feldstein:2006ce,Giudice:2019iwl,Dvali:2004tma,Dvali:2003br,Eroncel:2018dkg,Arkani-Hamed:2020yna,Csaki:2020zqz,Strumia:2020bdy,Cheung:2018xnu,Graham:2015cka,Nelson:2017cfv,Gupta:2018wif,TitoDAgnolo:2021pjo,TitoDAgnolo:2021nhd,Trifinopoulos:2022tfx,Cline:2018ebc,Kartvelishvili:2020thd,Khoury:2019ajl,Khoury:2021zao,Giudice:2021viw}. Besides the obvious appeal of actually finding the mechanism of interest, it can be argued that also not entirely successful attempts in this direction are of great benefit, as they might hint at generic features of dynamically selected Higgs parameters. This includes, of course, the initial observation of near-critical parameters for the Higgs potential. More curiously, two recently suggested, entirely independent mechanisms have proven capable of explaining the Higgs mass \textit{if} the instability scale $\mu_I$, i.e., the scale where the Higgs' quartic coupling vanishes, were significantly lowered just above the range of current observational constraints~\cite{Giudice:2021viw,Khoury:2019yoo,Khoury:2021zao,Kartvelishvili:2020thd,Khoury:2019ajl}. This would, for example, favor SM extensions including additional fermions with sizable Yukawa couplings to the Higgs and TeV-scale masses, such as particular realizations for neutrino masses~\cite{Pilaftsis:1991ug,Gluza:2002vs,Xing:2009in,He:2009ua,Ibarra:2010xw,Mitra:2011qr,Shaposhnikov:2006nn,Chauhan:2023pur}.\footnote{The prospect of the vacuum becoming destabilized at energies within the reach of future accelerators and astrophysical processes has also driven progress in the understanding of vacuum decay out of excited states more generally~\cite{Steingasser:2023gde,Steingasser:2024ikl}. Similarly, the need of such models to account for spatial variations in what were thought to be constants might lead to a renewed interest in soliton-like configurations and the discovery of new properties. See, e.g., \cite{Steingasser:2020lhj}.}

In these notes, I lay out a different approach. Rather than focusing on the development of concrete mechanisms for vacuum selection, I suggest identifying and investigating the consequences of generic features of near-criticality in particle physics and Cosmology. Most importantly, I argue that this amounts to also considering near-critical combinations of parameters when investigating Beyond-Standard-Model (BSM) models rather than \textit{only} focusing on natural combinations. While this approach is less predictive than the development of a concrete mechanism, it also comes with two major advantages. First, on a practical level, the development of concrete mechanisms is subject to many subtleties and uncertainties, and an improved intuition for the behavior of such models could prove as valuable guidance. More conceptually, disentangling predictions specific to one particular mechanism from more general properties may help with the interpretation of possible observations linked to near-criticality. 

These notes are structured as follows. First, in Sec.~\ref{sec:SM}, I review the near-criticality of the Standard Model (SM). This includes an alternative QPT of the SM potential, which plays a crucial role in the recently discovered \textit{metastability bound}, which I also review here. Next, in Sec.~\ref{sec:VacSel}, I review two recently proposed mechanisms for dynamical vacuum selection. These mechanisms not only serve as an important proof of concept for my underlying assumption, but are also crucial in identifying universal properties of dynamically selected vacua. Next, in Sec.~\ref{sec:MinEx}, I demonstrate in great detail how to identify and interpret near-critical combinations of parameters in a given BSM model.

\section{Higgs criticality in the Standard Model}\label{sec:SM}

In the SM, the (RG-improved, effective) potential of the Higgs is famously of the form
\begin{equation}\label{eq:VSM}
    V_{\rm SM, eff}(\mu ,H)= V_0 - \frac{1}{4}m_{\rm eff}^2 (\mu, H) H^2 + \frac{1}{4}\lambda_{\rm eff} (\mu, H) H^4.
\end{equation}
Remarkably, as reviewed throughout the remainder of this section, all three parameters in this potential appear fine-tuned. In the following, I first explain this behavior from the perspective of ``conventional'' particle physics, and then discuss an intriguing observation: The apparent tunings can be understood as the parameters in the potential being close to \textit{critical values} marking transitions between qualitatively different potentials, so-called \textit{quantum phase transitions} (QPTs)~\cite{Jaeger:1998,Carr2010,Sachdev2011}. 

\subsection{The hierarchy problem}\label{sec:Hierarchy}

Assuming the existence of \textit{some} physics beyond the Standard Model, the potential~\eqref{eq:VSM} needs to be understood as a \textit{(low-energy) effective potential}, and the Standard Model itself as a \textit{(low-energy) effective theory}, which is obtained from some more fundamental theory by integrating out heavier degrees of freedom~\cite{Brivio:2017vri}. Further assuming a natural choice of parameters, this often leads to a Higgs mass of order of the scale of new physics already at tree-level, e.g. in composite Higgs models or in scenarios in which the Higgs appears as the lightest mass eigenstate of a larger scalar sector with non-trivial mass matrix.

Moreover, even if the Higgs mass receives no corrections at tree level, it can be expected to do so at the loop level. To understand why, we can first consider this situation in the full theory, where the Higgs mass receives loop-level corrections from each particle it couples to. For a generic quantum field theory with a set of scalars $\Phi=\{H,S,... \}$, the one-loop correction to the effective potential is generally of the form~\cite{Markkanen:2018bfx,Markkanen:2018pdo,Mantziris:2022fuu}
\begin{equation}\label{eq:dV}
    \Delta V_{\rm 1-loop} = \frac{1}{64 \pi^2} \sum_{i \in {\rm particles}} n_i m_i(\Phi)^4 \left[ \log \left( \frac{m_i^2 (\Phi)}{\mu^2} \right) - d_i \right].
\end{equation}
Here, $n_i$ and $d_i$ are species-dependent numerical factors, while $m_i^2 (\Phi)$ is the i\textit{th} particle's (background-field dependent) mass \textit{at tree level}. As an example, for the SM Higgs we have $m_{H}^2= - \frac{m^2}{2}+ 3 \lambda H^2$. If $\phi$ takes its vacuum expectation value, $v^2=\frac{m^2}{2 \lambda}$, this simplifies to $m^2$, motivating the choice of convention in Eq.~\eqref{eq:VSM}. Based on dimensional arguments, it is easy to see that the tree-level mass of a generic particle is generally of the form
\begin{equation}\label{eq:m4exp}
    m_i^2 (\Phi)= \bar{m}_i^2 + \gamma_i H^2 + ..., \ \ {\rm s.t.} \ \ m_i^4 (\Phi)=\bar{m}_i^4 + 2 \gamma_i \bar{m}_i^2 H^2 + ... \ .
\end{equation}
This implies, in particular, that the Higgs mass receives quantum corrections of order $\delta m^2 \sim \frac{\gamma}{(4 \pi)^2} \bar{m}_i^2$ from each particle it is coupled to. Thus, if the Higgs were to couple to a particle with a bare mass $\bar{m}_i$, its mass could be expected to be roughly of order $\frac{\bar{m}_i}{4 \pi}$, unless the heavy particle's contribution to the effective mass is canceled by a fine-tuned tree-level Higgs mass.\footnote{A noteworthy exception from this reasoning is famously supersymmetry, which ensures the cancellation of each particle's contribution to the correction in Eq.~\eqref{eq:dV} through the introduction of a particle with an opposite contribution. In this case, however, the breaking of supersymmetry necessary to explain the absence of superpartners in low-energy experiments generally leads to a natural mass term not significantly lower than the breaking scale, and thus the lightest superpartners.}  Setting out from the observed value of the physical Higgs mass, $M_h=125$~GeV, this would suggest that signs of new physics should have appeared at scales roughly around $O(1-10)$~TeV, well within the range of LHC. This, however, has not happened, giving rise to the \textit{hierarchy problem}.

\begin{figure}[t!]
    \centering
    \includegraphics[width=0.6\textwidth]{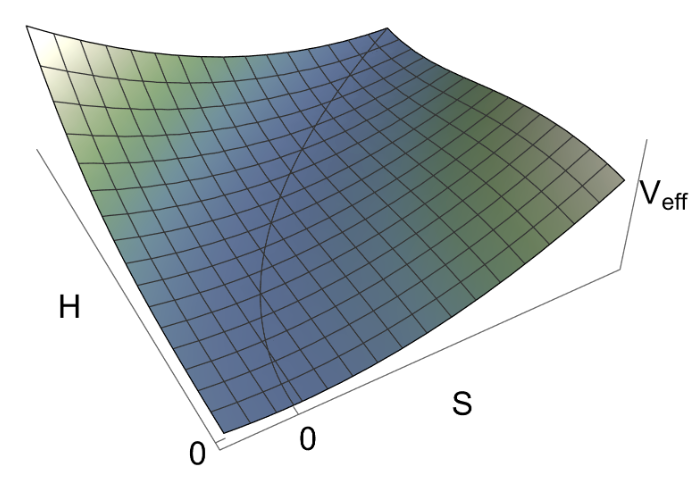}
    \caption{The potential of the Standard Model extended by an additional singlet $S$. For energies larger than the mass of $M$, the dynamics unfolds freely in field space. For smaller energies, it is confined to a ``valley'' approximately in $H$-direction for small field values. The degree of freedom of the effective theory can be understood as a displacement along this contour, which is highlighted in black.
    }
    \label{fig:EFTPot}
\end{figure} 

This problem may also be understood very efficiently within the framework of effective field theories. At energies smaller than the mass $M$ of some particle, the latter can be \textit{integrated out} to assure the validity of the perturbative treatment by avoiding large logarithms in the expansion. Formally, this amounts to replacing the full theory with an effective theory of the lighter degrees of freedom at scales below the \textit{matching scale} $\mu_M\sim M$~\cite{Georgi:1993mps,Kaplan:1995uv,Manohar:1996cq,Polchinski:1992ed,Pich:1998xt,Rothstein:2003mp,Skiba:2010xn,Burgess:2007pt}. On a semi-classical level, this procedure can be visualized easily for the extension of the SM by an additional scalar field, see Fig.~\ref{fig:EFTPot}. Most importantly for our purpose, appropriately capturing the properties of the full theory requires the parameters of the effective theory to be chosen s.t. the effective potential is continuous at the matching scale. In particular, retaining the information about the loop corrections induced by the heavy degree of freedom requires correcting the mass term by a so-called \textit{threshold correction},
\begin{equation}
	m_{\rm eff}^2 \sim m_{\rm UV}^2 \pm \frac{\gamma M^2}{(4 \pi)^2}.
\end{equation}
One would therefore generically expect the mass of the scalar to be comparable to that of the heavier degree of freedom up to one or two orders of magnitude at best.

The near-vanishing of the Higgs mass relative to the natural scale of whatever more fundamental theory underlies the Higgs sector can alternatively be understood in terms of a phase transition. Upholding the sign convention of Eq.~\eqref{fig:EFTPot}, the case $m^2>0$ corresponds to a potential with a non-trivial minimum $v$ with $v^2\sim m^2/\lambda$, and thus, spontaneous symmetry breaking. Varying the mass parameter towards the critical value $m_{\rm crit.}^2=0$ thus causes this minimum to approach the local maximum at the origin, with which it merges once the critical value is reached. In other words, the system undergoes a QPT from a symmetry-breaking phase to a symmetry-restored phase, with the mass parameter acting as an \textit{order parameter}~\cite{Jaeger:1998,Carr2010,Sachdev2011}.

\subsection{The metastability of the electroweak vacuum}

While the mass term of the Higgs dominates the shape of its potential at low energies, it becomes essentially negligible at high energies, where its effective potential can thus be brought to the simple form
\begin{equation}\label{eq:VHE}
    V_{\rm eff}(H) = \frac{1}{4}\lambda_{\rm eff} H^4.
\end{equation}
The value of the (effective) quartic coupling at high energies can, to leading order, be inferred from its beta function~\cite{Khoury:2021zao},
\begin{equation}\label{eq:betalambda}
    \beta_{\lambda}= (4 \pi)^{-2}\left[24 \lambda^2 - 6 y_t^4 + \frac{3}{8}\left( 2 g^4 +(g^2 +{g^\prime}^2)^2 \right) - \lambda \left( 9g^2 + 3 {g^\prime}^2 -12 y_t^2 \right) \right] + ... \ .
\end{equation}

\begin{figure}[t!]
    \centering
    \includegraphics[width=0.6\textwidth]{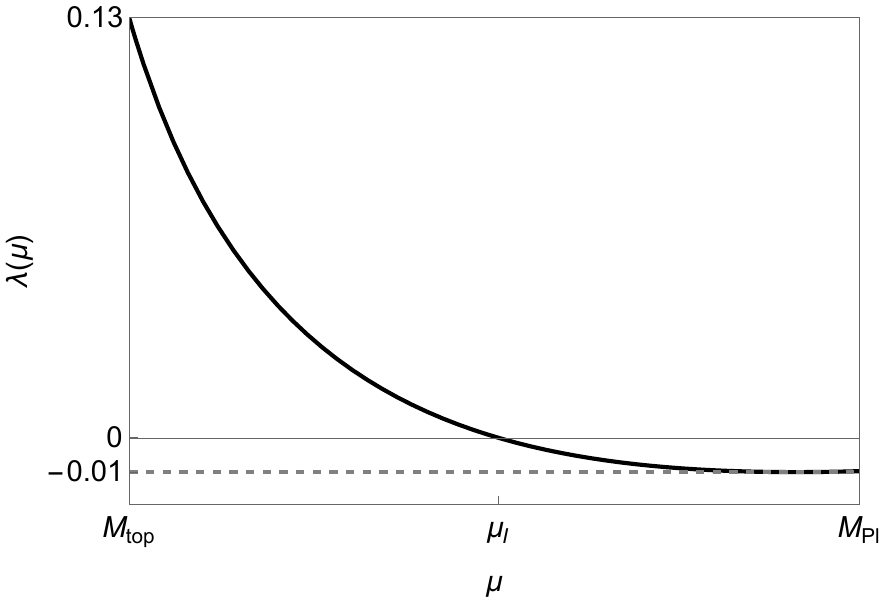}
    \caption{The running of the quartic coupling assuming the central values of the global PDG averages as inferred in Ref.~\cite{Huang:2020hdv}.
    }
    \label{fig:lambdarun}
\end{figure}

Using the most recent global averages of the particle data group as input for the values of the SM couplings near the top mass scale, integrating the SM beta functions at three-loop accuracy leads to the RG-trajectory of $\lambda$ shown in Fig.~\ref{fig:lambdarun}. For most of the input values allowed for by these data at $95\%$ confidence (see Ref.~\cite{Huang:2020hdv}), the quartic coupling turns negative at the so-called \textit{instability scale}. Around this scale the decline of the quartic slows down until $\lambda$ eventually reaches a minimum at $\mu_* \sim 10^{17}$~GeV (for the central values), causing it to essentially hover near a small, negative value up to the Planck scale. A full discussion of the observational error bars of these quantities can be found in Ref.~\cite{Steingasser:2023ugv}.

\begin{figure}[t!]
    \centering
    \includegraphics[width=0.9\textwidth]{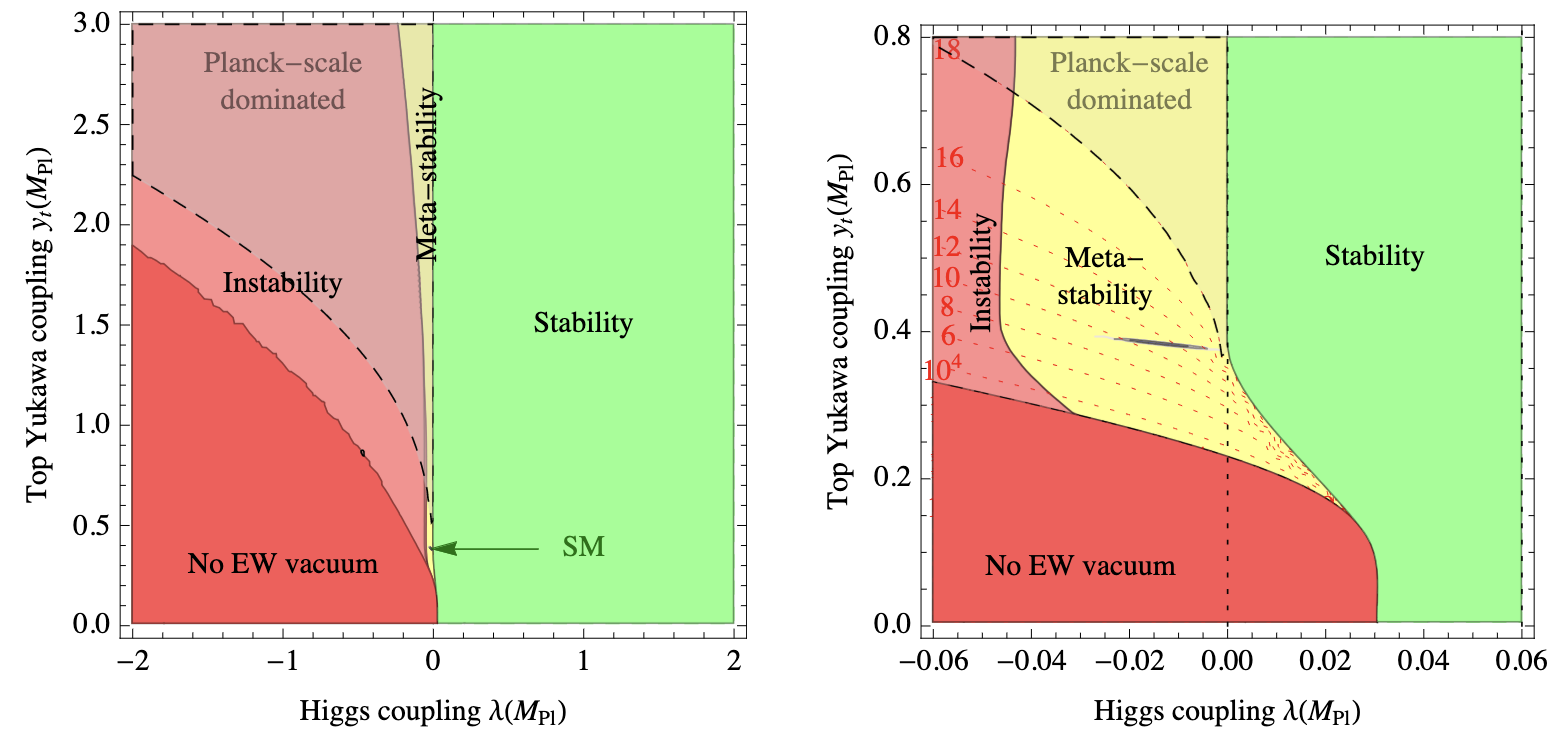}
    \caption{The stability of the electroweak vacuum as a function of the top Yukawa and quartic coupling at the Planck scale, as shown in Ref.~\cite{Buttazzo:2013uya}.}
    \label{fig:TPl}
\end{figure}

Before discussing the consequences of this behavior, it is worth understanding what causes it. From Eq.~\eqref{eq:betalambda}, it is easy to see that the running of $\lambda$ towards smaller values is driven by the contribution from the Yukawa coupling between the Higgs and the top quark, $y_t$. At low energies, this coupling takes relatively large values $y_t \sim 0.9$, s.t. its contribution to the overall beta function becomes dominant. At higher energies, $y_t$ itself is driven towards smaller values due to the QCD contribution $\propto g_s^3$ to its one-loop beta function. Thus, the negative term in the beta function becomes less effective, causing the quartic to turn back upwards. This behavior, and in particular the emergence of a near-vanishing quartic at the Planck scale, is remarkably sensitive to the values of the couplings near the top mass scale.\footnote{In principle, this could suggest simply adopting a top-down perspective. Assuming ongoing validity of the Standard Model, these ``fine-tuned'' parameters would be an inevitable consequence of the quartic taking \textit{some} perturbative value close to zero at the Planck scale. Throughout the remainder of this subsection, it should, however, become clear that the vacuum's stability is similarly sensitive to the coupling's Planck scale values, with the inferred metastability corresponding to a similarly special combination of parameters. This is made explicit, e.g., in Fig.~\eqref{fig:TPl} taken from Ref.~\cite{Buttazzo:2013uya}.}

It is evident from Eq.~\eqref{eq:VHE} that the quartic coupling turning negative implies that the potential becomes unbounded from below, while our vacuum is protected by a potential barrier just below $H \sim \mu_I$. This allows for the nucleation of a finite-sized bubble with field values beyond the barrier at its center, which subsequently expands with approximately the speed of light. This nucleation rate has been of great theoretical interest for decades, and only recently a complete calculation addressing all subtleties as well as taking into account gravitational corrections has been performed in Refs.~\cite{Andreassen:2017rzq,Khoury:2021zao,Steingasser:2022yqx,Chauhan:2023pur}. Most importantly, these works confirmed the existing leading-order result,
\begin{equation}
    \frac{\Gamma}{V} \sim \mu_S^4  e^{-\frac{8 \pi^2}{3 |\lambda (\mu_S)|} - ...  }.
\end{equation}
For the case of the pure Standard Model, the scale $\mu_S$ coincides with the scale $\mu_*$ where the quartic coupling reaches its minimum and the terms represented by the ellipsis vanish~\cite{Andreassen:2017rzq}. More generally, in particular taking into account leading-order contributions from gravity or higher-dimensional operators encoding BSM physics, $\mu_S$ is subject to a more complicated relation discussed in great details in Refs.~\cite{Khoury:2021zao,Chauhan:2023pur}, where $\mu_S$ was dubbed the \textit{instanton scale}.

For our purpose, the most important aspect of these results is their high sensitivity to the precise value of the quartic coupling, which can be illustrated through the error bars of the current prediction of the the lifetime. Again using the global averages given in Ref.~\cite{Huang:2020hdv} as input parameters and performing a full one-loop computation including the leading order gravitational effects, the lifetime of our vacuum can be determined as~\cite{Khoury:2021zao}
\begin{equation}
    \tau = 10^{983^{+1410}_{-430}}~{\rm years}.
\end{equation}
The significant error bar in the exponent, which was obtained by scanning over the $68\%$ error ellipsoid in the physical input parameters, is an immediate consequence of this result's strong sensitivity to the involved couplings.

An important subtlety is that this sensitivity is \textit{not} specific to the input parameters at low energies, whose uncertainty does indeed get amplified by the non-linearity of the running of the couplings over the roughly 15 orders of magnitude up to the relevant scale. Instead, the decay rate of the vacuum is also remarkably sensitive to the values of the relevant couplings near the Planck scale. See, e.g., Fig.~\ref{fig:TPl}, taken from Ref.~\cite{Buttazzo:2013uya}.\footnote{Note that, while these results remain true qualitatively, the precise regions shown in Fig.~\ref{fig:TPl} have changed. First, newer, more precise measurements have shrunk the error ellipsoid and moved it closer to the transition from stable to metastable regime. More importantly, these results did not take into account gravitational effects laid out in Refs.~\cite{Khoury:2021zao,Chauhan:2023pur}, which not only stabilize the vacuum directly, but also lower the relevant RG scale, shrinking the ``Planck-scale dominated'' region through their effect on the instanton scale.}

From the point of view of the tree-level potential, the destabilizing of the vacuum through the running of $\lambda$ and the ``subsequent'' tunneling can be understood as a QPT, as in this picture the change of the quartic through its running would be understood as the change of a physical parameter. When understanding the theory in terms of its full, RG-improved effective potential, this picture slightly changes. Since the running of the quartic is already incorporated into this potential, it would not actually correspond to a change of a parameter. Instead, one may consider the couplings' initial values at some particular reference scale, e.g., the Planck scale, as the theory's control parameter. This allows for a meaningful notion of a phase of the potential itself. For negative $\lambda (M_{\rm Pl})$, the phase of the potential is characterized by the existence of two minima, whereas positive values correspond to a potential with only one minimum. Interestingly, this idea of assigning a phase to the potential itself based on certain qualitative features - in this case, its number and type of minima - can also be applied to the hierarchy problem. 

From this perspective, the phase transition would be represented by a potential marking the transition point from a potential with two minima to a potential with one minimum. By considering the two phases of interest, it is easy to see that this transition would go along with the formation of an inflection point. A simple computation then shows that the location of this inflection point roughly coincides with the scale where the $\lambda$ reaches its minimum. More importantly from the perspective of near-criticality, this requires the non-field dependent part of the effective quartic coupling to be precisely $1/16$ of the coefficient in front of the terms proportional to $\ln(H/\mu)^2$~\cite{Bezrukov:2014bra}. This relation can hardly be explained through naturalness, and yet, lies very close to the boundary of the current $95\%$ error ellipsoid representing uncertainties in the relevant SM parameters. For this special set of parameters, one would find the inflection point just about an order of magnitude below the reduced Planck scale, suggesting that it might be of significance in Cosmology. And indeed, this observation gave rise to the idea of \textit{critical Higgs inflation}~\cite{Bezrukov:2009db,Bezrukov:2014ipa,Bezrukov:2014bra,Ezquiaga:2017fvi,Rubio:2018ogq}. Ordinary Higgs inflation, relying only on the potential $V(H)\sim \frac{\lambda}{4} H^4$, famously fails to produce the amplitude of the CMB fluctuations. This issue can, in principle, be cured through the introduction of a non-minimal coupling $\xi$ between the Higgs and gravity, 
which would allow for the correct amplitude if $\xi \sim 47,000 \cdot \sqrt{| \lambda |}$~\cite{Bezrukov:2007ep,Bezrukov:2008ej}. This requires either a huge value of $\xi$, or a very small value of $|\lambda|$. Interestingly, the latter possibility seems to be almost satisfied in the Standard Model due to its near-criticality. Moreover, the formation of the inflection point itself further flattens the potential in its vicinity, allowing for successful inflation with values of $\xi$ as small as $O(10)$~\cite{Bezrukov:2014bra,Ezquiaga:2017fvi,Rubio:2018ogq}.

\subsection{Cosmological constant problem}

Besides the field-dependent mass and self-interaction term, the Higgs potential can, in principle, also contain a constant term $V_0$. While this term is inconsequential in the context of particle physics, Einstein's equations imply that it would nevertheless manifest on cosmological scales. On this level, a constant term in the potential can be combined with the constant term $\Lambda_{\rm GR}$ in the gravitational action into a single background energy density,
\begin{equation}
	\rho_{\Lambda} = \left( V_0 + m_{\rm Pl}^2 \Lambda_{\rm RG} \right) ,
\end{equation}
where $m_{\rm Pl}^2$ is the \textit{reduced Planck mass}. The value of this combined parameter can be inferred in a straightforward manner as $\rho_{\Lambda} \sim (10^{-12}{\rm GeV})^4$~\cite{SupernovaCosmologyProject:1998vns,SupernovaSearchTeam:1998fmf}. Such a small number either requires all relevant contributions to be tiny in the first place or to cancel each other out almost perfectly. Moreover, just as the hierarchy problem and the metastability, such delicate cancellations would be highly sensitive to quantum corrections, which can be inferred directly from Eqs.~\eqref{eq:dV} and~\eqref{eq:m4exp}. Each of the field-independent terms in Eq.~\eqref{eq:m4exp} not only leads to a contributes $\propto \bar{m}_i^4$ to the total value of $V_0$, but also manifests as a term in the beta function of $V_0$ of the form $\beta_{V_0}\propto \bar{m}_i^4/(4 \pi)^2$~\cite{Markkanen:2018bfx}.

\subsection{An alternative phase transition - the metastability bound}\label{sec:MSBound}
So far, we have only considered those phase transitions of the Standard Model potential which naturally emerge if one assumes validity of the Standard Model up to very high energies. If, however, the instability scale $\mu_I$ were to be lowered drastically, e.g., through the effects of some additional light fermions with $O(1)$ Yukawa couplings, the Standard Model potential could actually lie close to another phase transition. 

\begin{figure}[t!]
    \centering
    \includegraphics[width=0.6\textwidth]{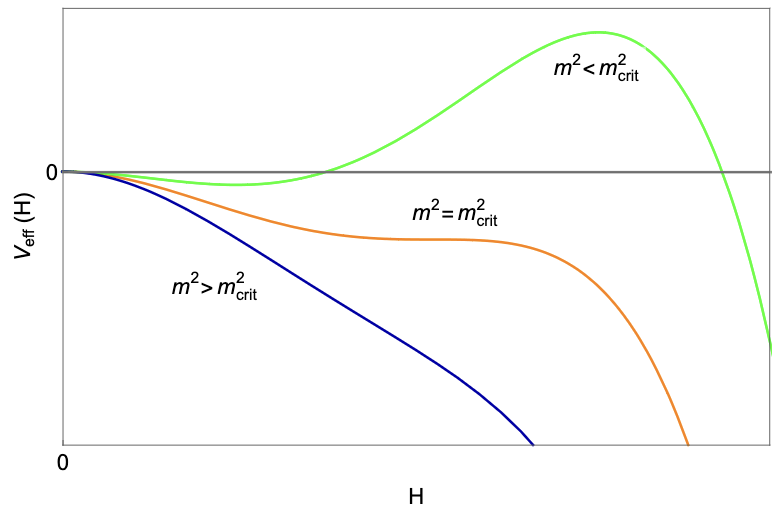}
    \caption{The potential for values of the mass parameter larger, smaller and identical to the critical value given by Eq.~\eqref{eq:mbound}.}
    \label{fig:MBound}
\end{figure} 

This phase transition can best be understood by setting out from a Standard Model-like potential with a non-trivial electroweak vacuum that ``flips'' at the instability scale $\mu_I$ due to the running of the quartic. The shape of this potential can be understood easily: For small values of the field, the negative mass term dominates, inducing a local maximum at the origin. Then, at some field value $v^2\sim m^2/\lambda$, the quartic term becomes dominant, causing the potential to bend upwards, giving rise to the electroweak vacuum. If now, however, $m_h^2 \gtrsim \mu_I^2$, this will only happen once the quartic coupling is already negative. This would prevent the formation of any minimum, see Fig.~\ref{fig:MBound}. More precisely, the phase transition occurs when the Higgs mass, instability scale and the quartic coupling's beta function $\beta_\lambda$ satisfy~\cite{Buttazzo:2013uya,Khoury:2021zao}
\begin{equation}\label{eq:mbound}
    m_h^2 = e^{-3/2} |\beta_\lambda (\mu_I)| \cdot \mu_I^2.
\end{equation}
What makes this relation conceptually interesting on its own is that it seems to relate the metastability of the vacuum to the Higgs mass. While this relation has been known for years~\cite{Buttazzo:2013uya}, its reconsideration from the perspective of near-criticality ultimately led to the discovery of the \textit{metastability bound} in Ref.~\cite{Khoury:2021zao}. This bounds states that in order for a metastable vacuum with the same symmetry breaking properties as in the SM to exist, the Higgs mass \textit{has} to be exponentially smaller than the scale $\Lambda_{\rm UV}$ where its sector is embedded into some more fundamental framework such as, e.g., a composite Higgs model. Relation~\eqref{eq:mbound} already provides an upper bound on the Higgs mass, so it only remains to show that $\Lambda_{\rm UV}^2 \gg \mu_I^2$. The argument can be found in full detail in Ref.~\cite{Khoury:2021zao} and heavily relies on recent insights into the calculation of the tunneling rate out of a false vacuum in a theory with classical scale invariance developed in Ref.~\cite{Andreassen:2017rzq}. In essence, these authors argue that quantum corrections break the classical scale invariance, and that the field values of the instanton (and thus, the relevant RG scale for all related calculations) are chosen by loop effects to be of order of the scale $\mu_*$ where the quartic coupling $\lambda$ reaches its minimum. This can be understood as the corresponding instanton giving rise to the dominant tunneling, as the Euclidean action of each instanton scales as $S_E \propto |\lambda (\mu)|^{-1}$, where the scale $\mu$ is linked to the typical field values within the instanton. As this dependence on the scale is driven by loop corrections, it is correspondingly suppressed. If now, on the other hand, the potential were to be extended by some dimension-six term, this would introduce a \textit{polynomial} dependence on the scale of interest. As such a correction would furthermore inevitably stabilize the vacuum, the dominant instanton would now be expected to be linked to some scale $\mu_S$ that is exponentially smaller than the scale of new physics $\Lambda_{\rm UV}$~\cite{Khoury:2021zao}. This implies that $\mu_S \ll \Lambda_{\rm UV}$. Lastly, as the leading-order contribution to the Euclidean action still scales as $|\lambda (\mu_S)|^{-1}$, achieving metastability with a not-too-large lifetime requires $\lambda$ to be sufficiently small at the scale $\mu_S$. As $\lambda$ only runs logarithmically, this implies $\mu_I \ll \mu_S$, again exponentially.

\section{Dynamical vacuum selection}\label{sec:VacSel}

The fact that all three parameters in the Higgs potential seem to give rise to problems related to fine-tuning can be understood as a hint that the properties of the Higgs sector are the result of mechanisms beyond the concepts of established particle physics, and the re-emerging pattern of near-criticality might suggest what kind of mechanisms to consider instead. It is well-known that near-criticality is a common feature of dynamical systems, where it can be achieved through a phenomenon called \textit{self-organized criticality}~\cite{PhysRevLett.59.381}. Simply speaking, critical points marking phase transitions often act as dynamical attractors, meaning that the system naturally evolves towards them. Under realistic conditions in which the system is subject to statistical fluctuations and only has a finite time to approach the critical point, this idea naturally leads to parameters very close, but not identical to the critical values. This behavior can now immediately be recognized in all three problems discussed in the previous section. It also distinguishes the dynamical approach from the conventional picture based on symmetries and gauge redundancies, which in many cases allows justifying the parameters of interest taking special values corresponding to a realization of the symmetry of interest. 

Altogether, this suggests that the open questions linked to the parameters of the Higgs sector could be naturally addressed by allowing them to be the result of some dynamical process. The conceptual simplicity of this idea is unfortunately contrasted by significant technical challenges affecting any attempt to actually realize it in a concrete model. Not only are there few precedents for such models to draw inspiration from, but there is also little evidence for the exact kind of model in which such a dynamics would have to unfold. In recent years, there has nevertheless been significant progress in this direction in the form of two independent mechanisms, \textit{self-organized localisation} and the idea of \textit{landscape statistics}. For the purpose of these notes, these approaches primarily serve as proofs of concept that the idea of a dynamical origin of the Higgs sector's parameter is in principle feasible and can be used to make predictions beyond the given parameters of the Higgs sector it is aimed to explain. They furthermore provide a good context to develop the tools and language to properly discuss these ideas. As an example, in the context of a landscape each set of Higgs parameters could be understood as linked to a different vacuum within the landscape, s.t. a dynamical selection of these parameters could be visualized as dynamically selecting one possible vacuum of some larger theory. This motivates referring to these ideas more generally as \textit{dynamical vacuum selection}. 

In the following, I very briefly review these two realizations of dynamical vacuum selection, with a strong focus on core conceptual ideas and apparently reoccurring behavior, which are important for the main point of these notes.

\subsection{Self-organized localisation}

\begin{figure}[t!]
    \centering
    \includegraphics[width=0.75\textwidth]{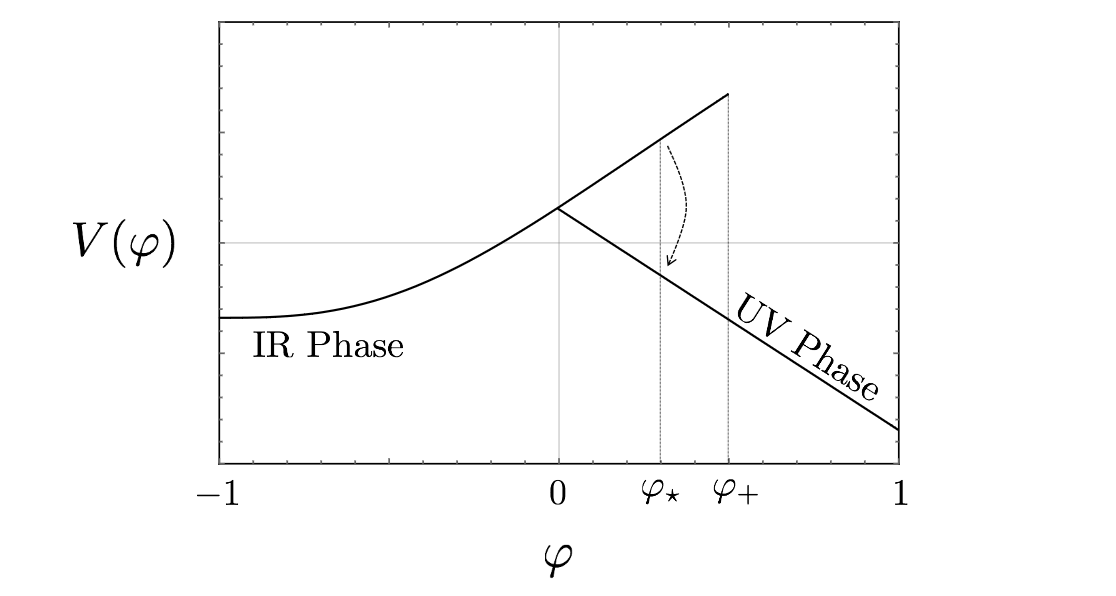}
    \caption{The backreaction on the apeiron potential caused by the Higgs' phase transition as shown in Ref.~\cite{Giudice:2021viw}.
    }
    \label{fig:Flip}
\end{figure}

Self-organized localisation (SOL) is based on the idea that the parameters of the Higgs potential are functions of some additional, light scalar field called \textit{apeiron}~\cite{Giudice:2021viw}. During inflation, this field is subject to large quantum fluctuations, requiring to describe its dynamics through a probability distribution $P_{\rm FP}(\phi,t)$ rather than a definite classical value. The dynamics of such a distribution can then be described through the well-known \textit{Fokker Planck equation}~\cite{Winitzki:1995pg,Vilenkin:1999kd}. Rewriting it in terms of the volume-weighted probability distribution $P$, this equation takes the form~\cite{Nakao:1988yi,Nambu:1988je,Nambu:1989uf,Linde:1993xx}
\begin{equation}\label{eq:FPV}
	\frac{\partial P}{\partial t} = \frac{\partial}{\partial \phi} \left[ \frac{\hbar}{8 \pi^2}  \frac{\partial (H^3 P)}{\partial \phi} + \frac{V^{\prime} P }{3 H}  \right] + 3H P,
\end{equation}
where $H$ is again the Hubble parameter. Setting out from this equation, the authors of Ref.~\cite{Giudice:2021viw} then show that, in a regime in which quantum fluctuations dominate over the classical dynamics, the volume-weighted probability distribution $P$ naturally develops a peak near the top of the potential for $\phi$. To relate this observation to the Higgs potential, the authors of Ref.~\cite{Giudice:2021viw} consider an effective potential of the form
\begin{equation}
	V(h,\phi)=- \frac{g_*^2}{4}\frac{\phi}{f} (h^2 - v^2)^2 + \omega(\phi).
\end{equation}
Here, $g_*$ is some coupling constant, $f$ is a constant of dimension energy, $v$ is the usual Higgs vev and $\omega$ is some suitable function of $\phi$. Let now $\phi$ start its time evolution from some generic negative value, corresponding to a positive effective Higgs quartic coupling $\lambda$. Assuming $\omega'(\phi)$ to be positive but sufficiently small for quantum fluctuations to dominate the dynamics, Eq.~\eqref{eq:FPV} immediately implies that $\phi$ will climb up the potential until it eventually reaches zero. Once it does, the sign of the effective quartic Higgs coupling changes. This, in particular, allows for the Higgs to evolve into a regime of large field values, where the existence of a true vacuum at $\langle h \rangle_{\rm UV}$ may be assumed. The Higgs settling into this vacuum causes a back-reaction on the potential for $\phi$ through the introduction of a new term,
\begin{equation}
	\Delta V \propto - \frac{g_*^2}{4} \cdot \frac{\langle h \rangle_{\rm UV}^4}{f} \cdot \phi.
\end{equation}
For a suitable choice of parameters, this effectively causes the potential to ``flip'', as illustrated in Fig.~\ref{fig:Flip} taken from Ref.~\cite{Giudice:2021viw}. Assuming the dynamics of $\phi$ to still be dominated by quantum fluctuations, these then drive $\phi$ back towards zero, possibly triggering the inverse transition, causing the dynamics to repeat. Altogether, this dynamics keeps the center of the probability distribution for $\phi$ localized near the top of the ``pyramid potential'' shown in Fig.~\ref{fig:Flip}, leading to a near-critical quartic coupling. A detailed discussion of how this picture can be used to infer the Higgs mass and cosmological constant is provided in Ref.~\cite{Giudice:2021viw}. For the purpose of these notes, the most important result obtained for these two parameters is the prediction of the Higgs mass, $m^2\sim |\beta_\lambda (\mu_I)| \cdot \mu_I^2$, where the \textit{instability scale} $\mu_I$ is the scale at which the quartic coupling turns negative. While this prediction is off by several orders of magnitude for the Standard Model, it can be significantly improved by lowering the instability scale through the effect of additional fermions together with some stabilizing term to keep the lifetime of the Universe above its current age.

\subsection{Landscape statistics}
An alternative approach towards a concrete mechanism of dynamical vacuum selection has been developed in Refs.~\cite{Khoury:2019yoo,Khoury:2019ajl,Kartvelishvili:2020thd,Khoury:2021zao} and, with slightly different assumptions and conclusions, Refs.~\cite{Khoury:2022ish,Khoury:2023ktz}. These works consider the possibility of a UV theory with numerous vacua, which together form a \textit{landscape}. The central idea is now to understand this landscape as a network, in which each node represents a vacuum. Two nodes are connected by a link if their corresponding vacua are connected by quantum tunneling. In this picture, a given observer living through a vacuum decay event could be understood as moving from one node another. Defining a measure through such an observer and considering only cases in which the overall probability distribution on the multiverse has not yet reached its equilibrium distribution, this allows to map the question of vacuum selection to one of search optimization. The observer can most likely be found in a region of the landscape in which the search algorithm determined by the tunneling rates between different vacua is most efficient. On the one hand, this excludes too stable vacua, which slow down the observers exploration of the landscape. A region of too short-lived vacua, on the other hand, bears the risk of the observer tunneling out of it entirely before having properly explored it. A more rigoros analysis including proper computations then ultimately allows for a concrete prediction of the typical lifetime for a vacuum within an optimal region. Remarkably, this ideal lifetime coincides with the Page time of the corresponding dS spacetime. Given the Hubble constant observed in our own vacuum, this prediction becomes
\begin{equation}
	\tau_{\rm ideal}\sim \tau_{\rm Page}\sim \frac{M_{\rm Pl}^2}{H_0^3} \simeq 10^{130}\ {\rm years},
\end{equation}
which is significantly smaller than the lifetime arising from the central values of the couplings as inferred in Ref.~\cite{Huang:2020hdv}. Furthermore, so far there seems to be no way of directly including the Higgs mass into this picture. Both of these issues can, however, be addressed through the results of Ref.~\cite{Khoury:2021zao}, and in particular the metastability bound discussed in Sec.~\ref{sec:MSBound},
\begin{equation}
	m_h^2 \lesssim e^{-3/2} |\beta_\lambda (\mu_I) \cdot \mu_I^2 \ll \Lambda_{\rm UV}^2,
\end{equation}
where the strength of the second inequality is controlled by the precise value of the desired lifetime and $\Lambda_{\rm UV}$ is the scale at which the Higgs sector is to be embedded into some underlying theory. As pointed out in the previous subsection, using even the most favorable set of couplings in compliance with the most recent input for the couplings would still lead to a bound multiple orders of magnitude above the observed value. However, just as for the Higgs mass prediction of SOL, this discrepancy can be lowered through a combination of additional fermions with TeV-scale masses to lower the instability scale and some additional new physics to stabilize the vacuum at higher energies.

More recently, the network picture has been revisited from the perspective of Bayesian statistics in Refs.~\cite{Khoury:2022ish,Khoury:2023ktz}. This perspective motivates a special choice of probability measure on the landscape, which can be shown to favor regions in the landscape that can be lie close to \textit{percolation criticality}. This means nothing more than a phase transition from a permeable (percolating) to an impermeable (non-percolating) state, and regions showcasing this behavior can be visualized as funnels. In the language of particle physics, this would amount to a series of vacua with near-degenerate vacuum energies. Interestingly, this prediction appears closely related to the independently conjectured \textit{Multiple Point Principle}, which assumes the existence of such a vacuum at the Planck scale~\cite{Froggatt:1995rt,Froggatt:2001pa,Das:2005ku,Froggatt:2005nb,Das:2005eb,Froggatt:2004nn,Froggatt:2006zc,Froggatt:2008hc,Das:2008an,Klinkhamer:2013sos,Froggatt:2014jza,Haba:2014sia,Haba:2014qca,Kawana:2014zxa,Hamada:2015fma,Hamada:2015ria,Haba:2016gqx,Haba:2017quk,McDowall:2018tdg,McDowall:2019knq,Maniatis:2020wfz,Kannike:2020qtw,Kawai:2021lam,Hamada:2014xka,Haba:2014zda,Kawana:2015tka,Hamada:2021jls,Racioppi:2021ynx,Huitu:2022fcw,Hamada:2015dja,Hamada:2017yji,Kawana:2016tkw,Ke:2024lel}. A similar feature also arises in the example for (near-)critical BSM physics discussed in Ref.~\cite{Steingasser:2023ugv}, which I review in Sec.~\ref{sec:MinEx}.

\section{Minimal example for BSM criticality}\label{sec:MinEx}

An important subtlety of the apparent tunings of the Standard Model is that their underlying conspiracies necessarily involve BSM physics. This is most evident for the hierarchy problem, which, from the EFT perspective laid out in Sec.~\ref{sec:Hierarchy}, is intrinsically a problem of the UV completion of interest - without BSM physics, there is no heavy particle that would contribute to the Higgs mass in the first place. Similarly, the cosmological constant would equally receive contributions from any BSM model, so that the necessary cancellations to obtain the observed value would naturally also have to involve BSM sectors. Lastly, should the inferred metastability persist, the previously discussed sensitivity of this property would also limit the size of contributions these BSM models can have on the running of $\lambda$. This suggests that the mechanism driving the Universe towards near-criticality should not only influence the Higgs sector. In other words, the assumption that there indeed exists some dynamical mechanism driving the parameters of the Higgs potential towards critical values also opens up the possibility that this mechanism affects unrelated BSM physics. 

\begin{figure}[t!]
    \centering
    \includegraphics[width=0.6\textwidth]{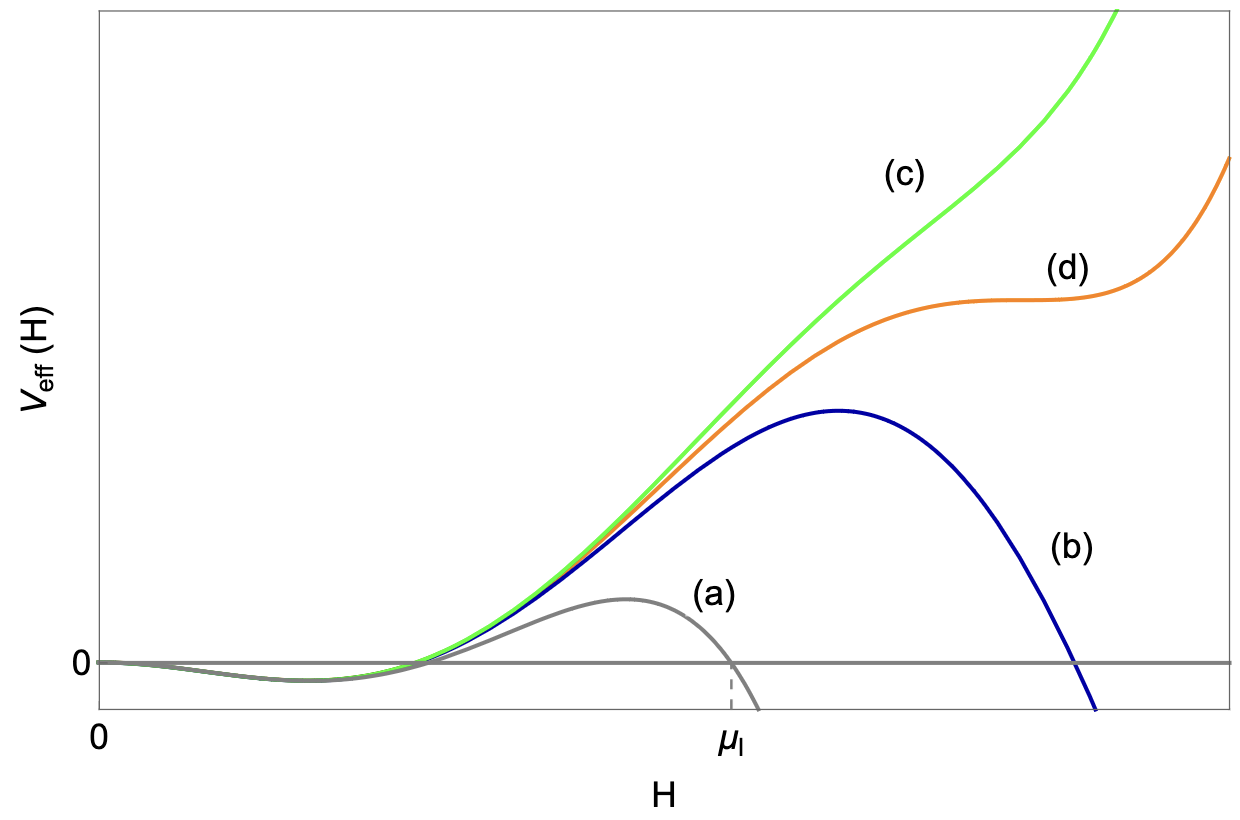}
    \caption{The Higgs potential extended by a dimension-six term, as in Eq.~\eqref{eq:VBSM}. For a vanishing order parameter, this potential is just the SM potential (a). For a value smaller than the critical value~\eqref{eq:C6crit}, a second minimum still forms, see case (b). For values larger than this critical value, however, no second extremum forms, see case (c). The potential (d) marks the transition between these two phases.
    }
    \label{fig:critdim6}
\end{figure}

A simple example for how this insight can be used to identify potentially interesting parts of parameters spaces of BSM models is the extension of the SM potential by an effective dimension-six operator considered in great detail in Ref.~\cite{Steingasser:2023ugv},
\begin{equation}\label{eq:VBSM}
    V_{\rm SM} \to V_{\rm SM} + \frac{C_6}{\Lambda_{\rm UV}^2}H^6.
\end{equation}
If the new physics encoded in this effective operator is indeed subject to the same tuning mechanism as the SM, we would expect to find the potential near a phase transition. To identify possible phase transitions, we first need to identify the possible phases of the potential~\eqref{eq:VBSM}. Considering the form of the unaltered Standard Model potential as given, the combination $C_6/\Lambda^2$ is the obvious choice for an order parameter of the full potential, s.t. different phases can be expected to correspond to different values of this parameter. If this parameter is large enough for the dimension-six term to become dominant at field values smaller than the instability scale, it completely stabilizes the potential. If it is not, the potential still bends downwards near the instability scale, and eventually reaches a minimum. Taking these two possibilities as distinct phases, it is straightforward to identify the critical value for $C_6/\Lambda^2$, which describes a potential with an inflection point marking the phase transition,
\begin{equation}\label{eq:C6crit}
    \left( \frac{C_6}{\Lambda^2} \right)_{\rm crit} =  \frac{1}{12 \sqrt{e}}  \cdot \frac{|\beta_\lambda|}{\mu_I^2}.
\end{equation}
This behavior is illustrated in Fig.~\eqref{fig:critdim6}.

Having identified the critical point, the next step would be to identify the near-critical region. While quantifying what exactly it means for the potential to be \textit{near} its critical configuration would indeed require a concrete vacuum selection mechanism, we can nevertheless identify distinct qualitative features. 

First, if the order parameter is slightly larger than its critical value~\eqref{eq:C6crit}, we expect to observe a near-inflection point. Potentials with such points have recently gained great interest in Cosmology, where they are in particular relied on for the production of Primordial Black Holes (PBHs)~\cite{Geller:2022nkr,Qin:2023lgo,Ezquiaga:2017fvi}. A well-understood subtlety of these mechanisms, however, is that they require a certain amount of fine-tunings not only to give rise to the necessary near-inflection points in the first place, but also to match inflationary observables and generate a sufficient population of PBHs. While this would pose a serious challenge in model building efforts based on the idea of naturalness, it would indeed appear natural if one were to rather look for near-critical models.

If, on the other hand, the order parameter were slightly smaller than the critical value~\eqref{eq:C6crit}, the inflection point would be split into a local maximum and local minimum. As this process happens continuously in $C_6/\Lambda^2$, small enough deviations from criticality would lead to a minimum with an energy of order of the inflection point, $V_{\rm min}\sim |\beta_\lambda(\mu_I)| \mu_I^4$. Most importantly, due to the nature of the critical point, this only holds true for order parameters that are fine-tuned against the critical value with an accuracy of a few percent~\cite{Steingasser:2023ugv}, whereas all natural values of this parameter lead to a second minimum with negative energy. In other words, both a near-inflection point and a second minimum with an energy larger than that of the electroweak vacuum would be a distinct sign for near-criticality of the full model including the dimension-six term.

The near-criticality of such a model could also be investigated on the particle physics frontier once a concrete model has been identified as the origin of the dimension-six term. This is demonstrated in great detail for the simplest example of an extension of the Standard Model by a singlet in Ref.~\cite{Steingasser:2023ugv}.

Remarkably, this again relates to the \textit{Multiple Point Principle}~\cite{Froggatt:1995rt,Froggatt:2001pa,Das:2005ku,Froggatt:2005nb,Das:2005eb,Froggatt:2004nn,Froggatt:2006zc,Froggatt:2008hc,Das:2008an,Klinkhamer:2013sos,Froggatt:2014jza,Haba:2014sia,Haba:2014qca,Kawana:2014zxa,Hamada:2015fma,Hamada:2015ria,Haba:2016gqx,Haba:2017quk,McDowall:2018tdg,McDowall:2019knq,Maniatis:2020wfz,Kannike:2020qtw,Kawai:2021lam,Hamada:2014xka,Haba:2014zda,Kawana:2015tka,Hamada:2021jls,Racioppi:2021ynx,Huitu:2022fcw,Hamada:2015dja,Hamada:2017yji,Kawana:2016tkw,Ke:2024lel}. While the exploration of this proposal has been very fruitful in general, the same can not be claimed about the question of why this principle should be realized in nature. This changes, however, once one is willing to accept the possibility of near-criticality of BSM physics. In our language, demanding the (approximate) validity of the MPP would be nothing else but assuming the theory to have parameters within a finite-sized subset of the near-critical part of parameter space.

\section{Conclusion}
The Higgs potential contains three parameters - the Higgs' mass and quartic coupling as well as a constant term. Each of these terms appears fine-tuned, manifesting in the hierarchy problem, the metastability of the electroweak vacuum, and the cosmological constant problem. This peculiar behavior seems at conflict with the principles conventional particle physics model building relies upon, but the details of these apparent tunings can be understood as hinting at a solution. All of these tunings manifest in parameters close to some \textit{critical values} marking the transition between qualitatively distinct classes, or \textit{phases}, of the Higgs potential. Such behavior is typical in dynamical systems, where such phase transitions often act as \textit{dynamical attractors}. In recent years, this has led to the development of two distinct classes of mechanisms capable of achieving this, \textit{multiverse statistics} and \textit{self-organized localisation}. First of all, these mechanisms serve as a \textit{proof of concept} that the idea of dynamically selecting the Higgs parameters is feasible and can lead to concrete predictions. They furthermore offer some level of guidance for the approach laid out in these notes: Given the many open problems complicating the construction of vacuum selection mechanisms, I suggest to identify universal features of near-critical models independent of the underlying mechanism, and search for their imprints in observations. The most important of these features are, of course, parameters close to critical values marking phase transitions. This translates to a surprisingly concrete procedure for model building. Once the general structure of the theory has been specified, one can first identify all possible phases of the model. Doing so allows to also identify all possible transitions between these phases, and with them their associated critical combinations of parameters and, finally, regions of parameter space corresponding to near-critical behavior. I demonstrated this procedure explicitly for the simplest possible Standard Model extension, in which near-criticality would manifest through the formation of near-inflection points or an additional minimum with very distinct energies. Depending on the precise values of the involved scales, both of these kinds of features could, in principle, lead to distinct imprints in future cosmological observations. Another important benefit of this approach is that it allows to revisit many models and problems from a fresh perspective. An important first example for a new result that is interesting independent from the idea of dynamical vacuum selection is the metastability bound. Thus, by further developing this radically new perspective we not only have a chance at discovering an entirely new kind of ``fundamental physics'', but also to learn more about the models we have already been studying for years, including the Standard Model.

\section*{Acknowledgments}
The parts of my own research which I discuss in these notes wouldn't have been possible without the generous support, excellent guidance and outstanding mentorship of both David. I. Kaiser and Justin Khoury. I also thank Gian Giudice for further inspiration and encouragement.

This work was made possible by the Walter Benjamin Programme of the Deutsche Forschungsgemeinschaft (DFG, German Research Foundation) -- 512630918. Portions of this work were conducted in MIT's Center for Theoretical Physics and partially supported by the U.S. Department of Energy under Contract No.~DE-SC0012567. This project was also supported in part by the Black Hole Initiative at Harvard University, with support from the Gordon and Betty Moore Foundation and the John Templeton Foundation. The opinions expressed in this publication are those of the author(s) and do not necessarily reflect the views of these Foundations.
\bibliographystyle{unsrtnat}
\bibliography{ref}

\end{document}